\documentclass[aps,pre,twocolumn,showpacs,superscriptaddress,10pt]{revtex4}
\usepackage{graphicx}
\usepackage{subfigure}
\usepackage{amsmath}
\usepackage{amsfonts}
\usepackage{amssymb}
\usepackage{amsthm}
\usepackage{times}
\usepackage[colorlinks,citecolor=blue,linkcolor=blue]{hyperref}
\usepackage{color}

\begin{document}

\title{Renormalized phonons in nonlinear lattices: A variational approach}

\newcommand{\Fudan}{\affiliation{State Key Laboratory of Surface
Physics and Department of Physics, Fudan University, Shanghai 200433,
China}}
\newcommand{\NUS}{\affiliation{Department of Physics and Centre for
Computational Science
and Engineering, National University of Singapore, 117546 Singapore}}
\newcommand{\NGS}{\affiliation{NUS Graduate School for Integrative Sciences and
Engineering, 117456 Singapore}}
\newcommand{\Tongji}{\affiliation{Center for Phononics and Thermal Energy
Science,
School of Physics Science and Engineering, Tongji University, Shanghai 200092,
China}}
\newcommand{\Graphene}{\affiliation{Graphene Research Centre, Faculty of Science, National
University of Singapore, 117542 Singapore}}
\newcommand{\Nanjing}{\affiliation{Collaborative Innovation Center of Advanced Microstructures, Nanjing University, Nanjing 210093, China}}

\author{Junjie Liu}
\Fudan
\author{Sha Liu}
\NUS
\author{Nianbei Li}
\Tongji
\author{Baowen Li}
\email{phylibw@nus.edu.sg}
\NUS
\Tongji
\NGS
\Graphene
\author{Changqin Wu}
\email{cqw@fudan.edu.cn}
\Fudan
\Nanjing

\newcommand{\fpuab}{FPU-$\alpha\beta$}
\newcommand{\fpub}{FPU-$\beta$}
\newcommand{\AP}{{r-ph}}
\newcommand{\QH}{quasi-harmonic}
\newcommand{\NFH}{NFH}
\newcommand{\GGB}{GGB}

\newcommand{\LH}{{LH}}
\newcommand{\UH}{{UH}}

\begin{abstract}

We propose a variational approach to study
renormalized phonons in momentum conserving nonlinear lattices with either symmetric or asymmetric potentials. To investigate the influence of pressure to phonon properties, we derive an inequality which provides both the lower and upper bound of the Gibbs free energy as the associated variational principle. This inequality is a direct extension to the Gibbs-Bogoliubov inequality. Taking the symmetry effect into account, the reference system for the variational approach is chosen to be harmonic with an asymmetric quadratic potential which contains variational parameters. We demonstrate the power of this approach by applying it to one dimensional nonlinear lattices with a symmetric or asymmetric Fermi-Pasta-Ulam type potential. For a system with a symmetric potential and zero pressure, we recover existing results. For other systems which beyond the scope of existing theories, including those having the symmetric potential and pressure, and those having the asymmetric potential with or without pressure, we also obtain accurate sound velocity.
\end{abstract}

\date{}


\maketitle

\section{Introduction}
Heat conduction in low dimensional anharmonic systems has attracted considerable interest in recent years \cite{Lepri.03.PR,Dhar.08.AP,Liu.12.EPJB}. Phonon, as the predominant heat carrier in insulating materials, undoubtedly lies in the heart of heat conduction. However, phonon bears a solid basis only in harmonic systems. The intrinsic nonlinearity in anharmonic systems will inevitably affect the behavior of phonon. Therefore, understanding the phonon properties in anharmonic systems represents an important question in the heat conduction field.

A renormalized phonon
(\AP{}) picture was then put forward independently by several groups using varying techniques
\cite{Alabiso.95.JSP,Lepri.98.PRE,Alabiso.01.JPA,Gershgorin.05.PRL,Li.06.EL,Gershgorin.07.PRE,
He.08.PRE,Li.13.PRE}. Within the scope of this picture, one can successfully interpret and understand a wide range of physical phenomena, including a theoretical description of the sound velocity \cite{Li.10.PRL} as well as scaling laws of thermal conductivity $\kappa(T)$ with temperature $T$ \cite{Li.07.EL,He.08.PRE}.

However, the generality of existing, state-of-the-art \QH{}
theories is limited. They were found to provide inaccurate predictions on the sound velocity of a nonlinear lattice with an asymmetric inter-particle potential \cite{Zhang.13.A}. Recently, a numerical method which aims to justify the validity of phonon concept in nonlinear lattices was proposed
\cite{Liu.14.PRB}. The existence of phonon modes in nonlinear lattices with asymmetric potentials is confirmed for a wide range of parameters. Hence, a unified \QH{} theory which extend to cover the general cases beyond harmonic is desirable.

In statistical mechanics, variational approaches are often used to obtain
approximate information of a nonlinear system via a reference system and an associated variational principle \cite{Girardeau.07.NULL}. The reference system whose properties can be easily obtained contains several variational parameters. In a variational scheme, an optimal reference system can be obtained by varying those parameters such that bounds of the variational principle go to a relative minimum or maximum.

In this paper, we develop a variational approach to study phonons in nonlinear lattices. We choose a general harmonic system as the reference system for this purpose and regard the so-obtained optimal harmonic system as an approximation to the original system from which we identify the properties of \AP{}s. We consider applications to one dimensional (1D) nonlinear lattices with either symmetric or asymmetric inter-particle Fermi-Pasta-Ulam (FPU) potentials \cite{Fermi.55.NULL} and demonstrate the power of our approach by comparing with molecule dynamics (MD) results.

This paper is organized as follows. In Sec.II, we introduce our variational approach. In Sec.III, we present numerical details.  In Sec.IV and Sec.V, we study nonlinear lattices with symmetric and asymmetric FPU potentials, respectively. Finally, we briefly summarize the work in Sec.VI.

\section{The variational approach}
\subsection{The variational principles}
Our current investigation aims to develop a \QH{} theory applicable for nonlinear lattices with asymmetric potentials. One of the attractive features of such systems is the existence of nonzero internal pressure due to the asymmetry of the potential. Then from a statistical mechanics point of view, it's convenient to use the language of the isothermal-isobaric ensemble which maintains constant temperature $T$, constant particle number $N$ and constant pressure $P$ (so-called $NPT$ ensemble) to describe them \cite{Brown.58.MP}. Their equilibrium thermodynamic properties can thus be well determined by the Gibbs free energy
\begin{equation}
G~=~-\beta^{-1}_T\ln\left[ \int e^{-\beta_T (H+PL)} d\mathbf{q}d\mathbf{p}\right],
\end{equation}
where $H$, $P$ and $L$ denote the Hamiltonian, pressure and volume (or length in 1D cases) of the system, respectively, $\beta_T\equiv 1/T$ is the inverse temperature (we set $k_B=1$), $\mathbf{q}$ and $\mathbf{p}$ are short for the products of all the coordinates and momenta of the system, respectively. The volume $L$ is a function of $\mathbf{q}$ only. Introducing
\begin{equation}
\mathcal{H}~=~H+PL,
\end{equation}
whose ensemble average is just the enthalpy of the system \cite{Landau.80.NULL}. The corresponding probability measure in phase space reads
\begin{equation}
\label{eq:npt-measure}
  \rho(\mathbf{q},\mathbf{p}) = \frac{e^{-\beta_T \mathcal{H}}}{\int e^{-\beta_T \mathcal{H}} d\mathbf{q}d\mathbf{p}}\equiv e^{-\beta_T (\mathcal{H}-G)}.
\end{equation}

A variational principle is an inequality satisfied by the physical quantity in which we may be interested \cite{Girardeau.07.NULL}. Hence we should look for inequalities for the Gibbs free energy $G$. To do this, we introduce a reference system with a Hamiltonian $H_0$ and prepare the original system and the reference system at same pressure. Then following this spirit, we integrate both sides of the following equality
\begin{equation}\label{eq:rho_e}
\rho(\mathbf{q},\mathbf{p})~=~\rho_0(\mathbf{q},\mathbf{p})e^{-\beta_T(\mathcal{H}-\mathcal{H}_0)-\beta_T(G_0-G)}
\end{equation}
over the whole phase space to yield
\begin{equation}
1~=~ e^{-\beta_T(G_0-G)}\cdot \left\langle
e^{-\beta_T(\mathcal{H}-\mathcal{H}_0)}\right\rangle_{\rho_0},
\end{equation}
where $G_0$ is the Gibbs free energy of the reference system and $\mathcal{H}_0=H_0+PL_0$, $\langle\cdot\rangle_{\rho_0}$ denotes the ensemble average under the probability measure
$\rho_0(\mathbf{q},\mathbf{p})\equiv e^{-\beta_T(\mathcal{H}_0-G_0)}$.

Taking logarithm
over both sides of the above equation and using the Jensen's inequality for exponential functions, i.e., $\langle e^{x}\rangle\geqslant e^{\langle x\rangle}$(following
\cite{Decoster.04.JPA}), we have
\begin{equation}
\label{eq:up}
G~\leq~G_0+\langle \mathcal{H}-\mathcal{H}_0\rangle_{\rho_0}.
\end{equation}
By switching the role of the nonlinear system and the reference system in Eq. (\ref{eq:rho_e}), we obtain
\begin{equation}
\label{eq:low}
G~\geq~G_0+\langle \mathcal{H}-\mathcal{H}_0\rangle_{\rho}
\end{equation}
with $\langle\cdot\rangle_{\rho}$ the ensemble average with respect to the probability measure $\rho(\mathbf{q},\mathbf{p})$ [c.f. Eq. (\ref{eq:npt-measure})]. These two inequalities [Eqs. (\ref{eq:up}) and (\ref{eq:low})] give the upper and lower bound of $G$, respectively.

If the pressure vanishes, i.e., $P=0$, the Gibbs free energies $G$ and $G_0$ reduce to the Helmholtz free energies $F$ and $F_0$, respectively, the enthalpy goes to the energy, the probability measure Eq. (\ref{eq:npt-measure}) should also be replaced by the canonical measure
\begin{equation}\label{eq:ca_dis}
  \rho^c(\mathbf{q},\mathbf{p}) = \frac{e^{-\beta_T H}}{\int e^{-\beta_T H} d\mathbf{q}d\mathbf{p}}\equiv e^{-\beta_T (H-F)}.
\end{equation}
Then we found that the inequality Eq. (\ref{eq:up}) recovers the well known Gibbs-Bogoliubov (GB) inequality \cite{Gibbs.10.NULL}
\begin{equation}
\label{eq:gb}
  F\le F_0+ \left\langle H-H_0\right\rangle_{\rho_0^c},
\end{equation}
where $F_0$ is the Helmholtz free energy of the reference system and $\langle\cdot\rangle_{\rho_0^c}$ stands for the ensemble average with respect to the canonical measure $\rho_0^c(\mathbf{q},\mathbf{p}) ~[\equiv e^{-\beta_T (H_0-F_0)}]$. And the inequality Eq. (\ref{eq:low}) goes to
\begin{equation}\label{eq:gb_lower}
  F\ge F_0+ \left\langle H-H_0\right\rangle_{\rho^c}
\end{equation}
with $\langle\cdot\rangle_{\rho^c}$ the ensemble average with respect to the probability measure $\rho^c(\mathbf{q},\mathbf{p})$ [c.f. Eq. (\ref{eq:ca_dis})]. As a lower bound of the Helmholtz free energy $F$ \cite{Feynman.82.NULL,Girardeau.07.NULL}, it's little applied since the ensemble average in it can not be evaluated analytically. However, it is more accurate than the upper bound in determining the free energy of solids \cite{Morris.95.PRL,Barnes.02.JCP}. As can be seen later, in our case for determining the sound velocity for anharmonic lattices, it is still the lower bound that gives better prediction comparing to the upper bound.

\subsection{The nonlinear systems}
We consider 1D momentum conserving nonlinear lattices described by the
general Hamiltonian \cite{Li.12.RMP}
\begin{equation}
H~=~\sum_{n=1}^{N}\left[\frac{p_n^2}{2m}+V(q_n-q_{n-1})\right],
\end{equation}
where $N$ is the particle number, $p_n$ denotes the momentum of $n$-th particle, $q_n=x_n- n r$
denotes the displacement of $n$-th particle from its equilibrium position $nr$ with $x_n$ the absolute position and $r$ the equilibrium distance for
the interaction bond,
and $V$ represents the inter-particle potential. For brevity and without loss of generality, we take $m=1$ and $r=1$ as the unit of mass and length, respectively.

The average lattice spacing $a$ is then given by $1+\langle q_n-q_{n-1}\rangle_{\rho}$, and the average lattice length $\bar{L}\equiv\langle L\rangle_{\rho}$ equals $Na$ for an $N$-particle lattice. We say that the lattice is at
its natural length if $a$ equals $r$ [=1], namely, $\bar{L}=N$ for an $N$-particle lattice. The average length can be changed to other values by applying pressure.

Furthermore, we introduce $\delta_n\equiv q_n-q_{n-1}$.
If the potential satisfies $V(\delta_n)=V(-\delta_n)$, we refer it
to a symmetric potential, otherwise we call it an asymmetric one. In terms of $\delta_n$ and $p_n$, the equations
of motion (EOMs) read
\begin{eqnarray}
\dot{\delta}_n &=& p_{n+1}-p_n,\\
\dot{p}_n &=& V'(\delta_n)-V'(\delta_{n-1}),
\end{eqnarray}
where the dot and the prime denote the time and space derivative, respectively.

Specifically, in this work, we focus on the FPU lattices with the inter-particle potential \cite{Fermi.55.NULL}
\begin{equation}\label{eq:abp}
V(\delta_n)~=~\frac{1}{2}\delta_n^2+\frac{\alpha}{3}\delta_n^3+\frac{\beta}{4}
\delta_n^4,
\end{equation}
which has become an archetype 1D nonlinear system in statistical mechanics \cite{Ford.92.PR,Berman.05.C}. We call the lattice with $\alpha= 0$ the \fpub{} lattice. Otherwise, we refer it to the
\fpuab{} lattice. For simplicity but without loss of generality, we only study the case of $\beta=1$ in this paper. It is evident that the \fpub{} lattice has a symmetric potential while the \fpuab{} lattice has an asymmetric one. Therefore, an \fpub{} lattice with natural length has zero pressure, while an \fpuab{} lattice with natural length has a nonvanishing pressure due
to the asymmetry of the potential. Nevertheless, our variational principles [Eqs. (\ref{eq:up}) and (\ref{eq:low})] enable us to study them within the same theoretical scheme.

\subsection{The harmonic reference system $H_0$}
Phonons bear a solid basis only in harmonic systems, in order to develop an effective phonon theory for nonlinear lattices, we should choose a quadratic $H_0$ as a reference system. Moreover, the present work investigates nonlinear lattices with asymmetric inter-particle potentials. Taking those into consideration, we consider a harmonic reference system in the following form
\begin{equation}\label{eq:h0}
H_0~=~\sum_{n=1}^{N}\left[\frac{p_n^2}{2}+V_0(\delta_n)\right];\quad
V_0(\delta_n) ~=~\frac{K}{2}(\delta_n-d)^2,
\end{equation}
in which $K$ and $d$ are variational parameters with $K$ being the effective elastic constant and $d$ quantifying the degree of asymmetry of the
potentials. Note that by setting $d=0$, we can use $H_0$ to deal with nonlinear systems with symmetric potentials as well.

The dispersion relation of this reference system is given by
\begin{equation}\label{ome}
\omega_k~=~2\sqrt{K}\sin\frac{ka}{2},
\end{equation}
where $k$ is the wavenumber. It can be
regarded as an approximate dispersion relation for the \AP{}s in nonlinear lattices \cite{Liu.14.PRB}. The corresponding sound velocity reads
\begin{equation}\label{eq:vs}
 c_s= \sqrt{K}a.
\end{equation}

The variational method is
then to select the optimal reference systems with parameterized Hamiltonian $H_0$ that
either minimize the right hand side (r.h.s.) of
Eq. (\ref{eq:up}) or maximize the r.h.s. of Eq. (\ref{eq:low}). This strategy then provides approximations
for the Gibbs free energy and allows us to find optimal $H_0$ for the system $H$.
The process is going to be detailed
in the next subsection.

\subsection{Determining the optimal harmonic systems}
Note that $dq_1dq_2\cdots dq_N=d\delta_1d
\delta_2\cdots d\delta_N$ in the large $N$ limit (the Jacobian is unity), then for a system with the probability measure Eq. (\ref{eq:npt-measure}), we have
\begin{equation}
\rho(\mathbf{q},\mathbf{p})=\prod_{n=1}^N\rho_s(\delta_n,p_n),
\end{equation}
where $\rho_s$ denotes the marginal phase space distribution for the single site variables $\delta_n$ and $p_n$ of $n$-th particle \cite{Dugdale.54.PR,Spohn.14.JSP}
\begin{equation}
\label{eq:single_rho}
  \rho_s(\delta,p)= \frac{1}{z} \exp\left\{-\beta_T\left[\frac{p^2}{2}+ V(\delta)+P\delta\right]\right\}
\end{equation}
with $z$ the corresponding partition function
\begin{equation}\label{eq:z}
  z= \iint \exp\left\{-\beta_T\left[\frac{p^2}{2}+ V(\delta)+P\delta\right]\right\} d\delta d p.
\end{equation}
Such a result can be readily tested
using MD simulations, see Sec. \ref{sec:md}.

For a system with sufficiently large particle number $N$,
the Gibbs free energy can be expressed as
\begin{equation}\label{eq:g_exact}
  G= Ng= -N\beta_T^{-1} \ln z,
\end{equation}
where $g= -\beta_T^{-1}\ln z$ is the Gibbs free energy per particle. So the Gibbs free energy of the reference system $H_0$ reads
\begin{equation}\label{eq:g0}
  G_0 ~=~ - N\beta_T^{-1} \ln z_0
\end{equation}
with $z_0$ the partition function determined by Eqs. (\ref{eq:h0}) and (\ref{eq:z}). It can be evaluated analytically, which gives
\begin{equation}
z_0~=~\frac{2\pi}{\beta_T \sqrt{K}}\mathrm{exp}\left[-\beta_T\left(Pd-\frac{P^2}{2K}\right)\right].
\end{equation}

Using the single-site distribution Eq. (\ref{eq:single_rho}), and noticing that the length $L$ and $L_0$ have a same expression, namely, $\sum_n(1+\delta_n)$, the averages in Eqs. (\ref{eq:up}) and (\ref{eq:low}) read
\begin{eqnarray}
\langle \mathcal{H}-\mathcal{H}_0\rangle_{\rho_0} &=& N\cdot\left\langle
V(\delta)-V_0(\delta)\right\rangle_{\rho_s^0}; \\
\langle \mathcal{H}-\mathcal{H}_0\rangle_{\rho} &=& N\cdot\left\langle
V(\delta)-V_0(\delta)\right\rangle_{\rho_s},
\end{eqnarray}
where $\left<\cdot\right>_{\rho_s^0}$ and $\left<\cdot\right>_{\rho_s}$ denote single-site averages with respect to the probability measure Eq. (\ref{eq:single_rho}) with the potential being $V_0$ and $V$, respectively. We will use these symbols in the rest of the paper.

\subsubsection{\upshape{The upper bound harmonic system}}
Firstly, we focus on the upper bound of $G$ [Eq. (\ref{eq:up})].  Minimizing it with respect to the variational parameters $K$ and $d$, i.e.,
\begin{eqnarray}
\frac{\partial}{\partial K}[G_0+\langle \mathcal{H}-\mathcal{H}_0\rangle_{\rho_0}] &=& 0,\\
\frac{\partial}{\partial d}[G_0+\langle \mathcal{H}-\mathcal{H}_0\rangle_{\rho_0}] &=& 0,
\end{eqnarray}
we obtain the following coupled equations:
\begin{eqnarray}
d_U &=& \frac{\left\langle(V-V_0)(\delta +\frac{P}{K_U})\right\rangle_{\rho_s^0}}{\left\langle V-V_0\right\rangle_{\rho_s^0}}, \label{eq:d2}\\
K_U &=&
\frac{\left\langle V-V_0\right\rangle_{\rho_s^0}}{\beta_T\left\langle(V-V_0)(\delta-d_U+\frac{P}{K_U}
)^2\right\rangle_{\rho_s^0}}.\label{eq:k2}
\end{eqnarray}
It can be readily tested that the second order derivatives are positive at $(d_U, K_U)$ so that this set of solution indeed gives the minimum of the upper bound. Thus $K_U$ and $d_U$ determine the optimal upper bound of the Gibbs free energy $G$ and the corresponding optimal Harmonic reference system (denoted as the \UH{} system).

\subsubsection{\upshape{The lower bound harmonic system}}
Now we turn to the lower bound of $G$ [Eq. (\ref{eq:low})]. Similarly, we maximize it with respect to the variational parameters $K$ and $d$, the optimal solution reads
\begin{eqnarray}
d_L &=& \langle \delta\rangle_{\rho_s}+\frac{P}{K_L},\label{eq:d1} \\
K_L &=& \frac{1}{\beta_T \left[\langle \delta^2\rangle_{\rho_s}-(\langle
\delta\rangle_{\rho_s})^2\right]}.\label{eq:k1}
\end{eqnarray}
We have also checked that $K_L$ and $d_L$ indeed give the maximum of the lower bound, thus them uniquely determine the optimal lower bound of the Gibbs free energy $G$ and also
the corresponding optimal harmonic reference system (denoted as the \LH{} system).
Note that $K_L$ does not rely on $d_L$ and is completely determined by Eq. (\ref{eq:k1}).

The \LH{} system and the \UH{} system can be regarded as effective descriptions of the original nonlinear system. If we use these two optimal systems to construct the effective theory of \AP{}s in the original nonlinear system, then the dispersion relation should follow Eq. (\ref{ome}) with $K$ given by $K_L$ or $K_U$. The accuracy of the results can be evaluated by comparing the calculated sound velocity with the predictions using Eq. (\ref{eq:vs}). In the following, we present numerical details and then apply this strategy to two 1D nonlinear lattices, namely, the \fpub{} lattice and the \fpuab{} lattice.

\section{Numerical Details \label{sec:md}}
\begin{figure}[tbh]
\includegraphics[width=0.85\columnwidth]{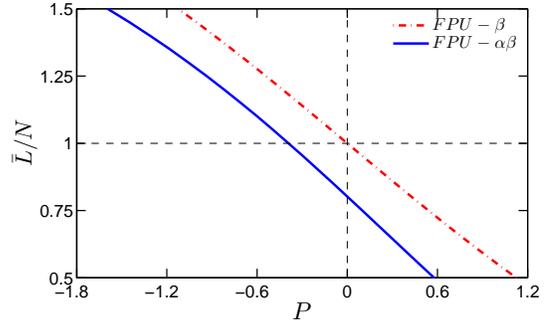}
\caption{Average lattice length $\bar{L}$ as a function of pressure $P$. The dashed-dotted line indicates that for the 1D \fpub{} lattice with $T=1$. The solid line denotes that for the 1D \fpuab{} lattice with $T=1$ and $\alpha=1$.}
\label{fig:pl}
\end{figure}

In this work, we consider systems either with or without pressure $P$. The average lattice length $\bar{L}$ can be controlled by adjusting the pressure, as demonstrated in Fig. \ref{fig:pl}.
It can be clearly seen that for the \fpub{} lattice once the pressure is nonzero, the average length of the lattice is away from the natural length, i.e., $\bar{L}\neq N$. It further indicates that we can apply a positive or negative pressure to this lattice system in order to compress or elongate its total length. But for systems with asymmetric potentials, i.e., the \fpuab{} lattice, the pressure would be nonzero at its natural length. Note that the average length $\bar{L}$ equals $N(1+\langle\delta\rangle_{\rho_s})$, so a stressless system with an asymmetric potential corresponds to an average length $N(1+\langle\delta\rangle_{\rho_s^c})$, where $\rho_s^c$ means the canonical single-site distribution, i.e., $\rho_s$ [Eq. (\ref{eq:single_rho})] with $P=0$. Particularly, for the \fpuab{} lattice with zero pressure, the average length is smaller than $N$ as can be seen from the figure.

An $NPT$ system with pressure $P$ can be prepared either by applying an external pressure $P$ to the end particles with free boundary condition or by fixing the total length $L$ to $\bar{L}(P)$, where $\bar{L}(P)$ as a function of $P$ is just depicted in Fig. \ref{fig:pl}. In the present work we adopt the second approach. Specifically, a modified periodic boundary condition is thus used to fix the total length at a certain value, namely, $q_N-q_0= \bar{L}- N=N\langle\delta\rangle_{\rho_s}$. For such systems, a symplectic integrator $\mathrm{SABA_2}$
with a corrector $\mathrm{SABA_2C}$ \cite{Laskar.01.CMDA} is adopted to integrate the EOMs
with a time step $h=0.02$ to ensures the conservation of the total system energy and momentum
up to high accuracy $10^{-14}$ in the whole run time, a transient time of order $10^7$ is used
to equilibrate the system.

\subsection{Single-site distributions}
We first check the single-site distributions in Eq. (\ref{eq:single_rho}). After the system reaches its thermal equilibrium state,
we can obtain time series of $v_n(\text{equals}~p_n)$ and $\delta_n$ of the $n$-th particle
($n$ can be an arbitrary integer from 1 to $N$). We then plot their histograms and compare the envelopes against $\rho_s=\rho_v\rho_{\delta}$ with
\begin{eqnarray}
\rho_v &=& \frac{1}{\sqrt{2\pi}}\mathrm{exp}\left(-\beta_T\frac{v^2}{2}\right),\label{eq:single_v}\\
\rho_{\delta} &=& z_{\delta}^{-1}\mathrm{exp}\left[-\beta_T\left(\frac{1}{2}\delta^2+\frac{\alpha}{3}
\delta^3+\frac{1}{4}\delta^4+P\delta\right)\right],\label{eq:single_q}
\end{eqnarray}
where $z_{\delta}=\int d\delta\, \mathrm{exp}[-\beta_T(\frac{1}{2}\delta^2+\frac{\alpha}{3}
\delta^3+\frac{1}{4}\delta^4+P\delta)]$.

\begin{figure}[tbh]
\includegraphics[width=0.9\columnwidth]{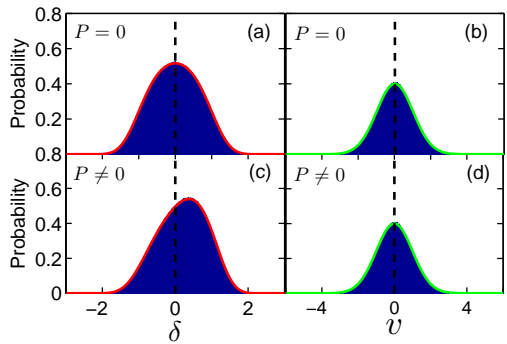}
\caption{Single-site distributions for the 1D \fpub{} lattice with $P=0$ (upper panel) or $P\neq 0$ (lower panel). The blue parts in the left and right column denote the histogram of $\delta$ and $v$, respectively. The
green and red solid lines are $\rho_{v}$ and $\rho_{\delta}$ [c.f., Eqs. (\ref{eq:single_v}) and (\ref{eq:single_q})], respectively.}
\label{fig:fpudis}
\end{figure}

\begin{figure}[tbh]
\includegraphics[width=0.9\columnwidth]{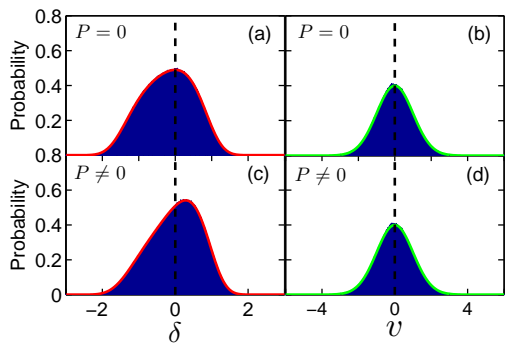}
\caption{Single-site distributions for the 1D \fpuab{} lattice with with $P=0$ (upper panel) or $P\neq 0$ (lower panel). The blue parts in the left and right column denote the histogram of $\delta$ and $v$, respectively. The
green and red solid lines are $\rho_{v}$ and $\rho_{\delta}$ [c.f., Eqs. (\ref{eq:single_v}) and (\ref{eq:single_q})], respectively.}
\label{fig:afpudis}
\end{figure}

To illustrate the comparison, we simulate the \fpub{} lattice with $N=2048$, $T=1$, and $\bar{L}=N$ or $\bar{L}=1.2N$, the corresponding pressure is 0 or -0.4309 according to Fig. \ref{fig:pl}, respectively. A similar simulation is carried out on the \fpuab{} lattice with $N=2048$, $T=1$, $\alpha=1$, and $\bar{L}=0.8023N$ or $\bar{L}=N$, with the pressure equals 0 or -0.3904, respectively. The results are depicted in Figs. \ref{fig:fpudis} and \ref{fig:afpudis}. $\rho_v$ and $\rho_{\delta}$ are shown as green and red solid curves in the figures, respectively. Perfect agreements between theoretical
curves and the MD results are presented, which indicates the validity of the single-site distribution in Eq. (\ref{eq:single_rho}).

\subsection{Sound velocity}
\begin{figure}[tbh]
\includegraphics[width=0.9\columnwidth]{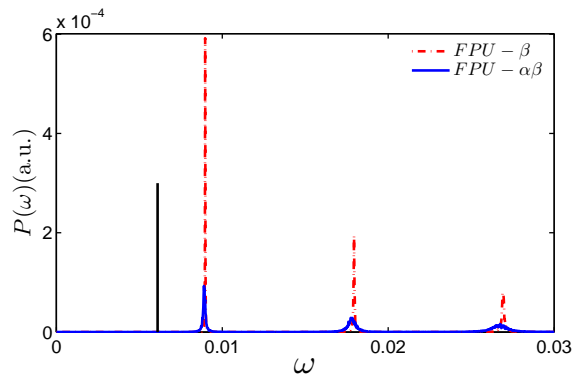}
\caption{Power spectrum $P(\omega)$ in the low frequency regime. The dashed-dotted line stands for the \fpub{} lattice at its natural length. The solid line represents the \fpuab{} lattice with $\alpha=1$ and natural length. For comparison, the lowest harmonic frequency is depicted as black solid line.}
\label{fig:ps}
\end{figure}

Now we turn to the calculation of the sound velocity. So far, there are mainly three numerical methods which can obtain the sound velocity of the nonlinear lattices. The first method regards the moving velocity of front peaks of the equilibrium spatiotemporal correlation function as the sound velocity \cite{Li.10.PRL,Zhao.06.PRL}. However, a broadening of front peaks at high temperature or strong anharmonicity will affect its accuracy. The second one is to look at the frequency of the lowest phonon peak of the power spectrum
of an $N$-particle lattice \cite{Zhang.13.A}
\begin{equation}
\omega_1~=~2\frac{c_s}{a}\sin\frac{\pi}{N},
\end{equation}
where $c_s$ is the sound velocity. This lowest phonon peak can always be detected in the power spectrum with high resolution (see Fig. \ref{fig:ps}), thus the sound velocity can be well determined by $\omega_1$. The third one obtains the sound velocity from the dispersion relation \cite{Liu.14.PRB}. This method, although follows the definition of the sound velocity, is most computationally expensive comparing with the other two, because the calculation of the dispersion relation is time consuming.

Therefore, in this study, we choose the second method to obtain MD results of the sound velocity of nonlinear lattices. According to the Wiener-Khinchin theorem \cite{Chatfield.89.NULL}, the power spectrum $P(\omega)$ can be obtained by doing Fourier transform of the velocity autocorrelation of a single particle $\langle v_n(t)v_n(0)\rangle$ with $t=0,h,2h,\cdots,t_{max}$. In order to get smooth phonon peaks in the power spectrum, the maximum correlation time $t_{max}$ whose inverse $1/t_{max}$ determines the frequency resolution is set to be $2^{24} h$ under the condition that $N=1024$. Fig. \ref{fig:ps} presents the power spectrum of the \fpub{} lattice and the \fpuab{} lattice with $\alpha=1$ at their natural length. $T=1$ is used for both models. It is apparent that phonon peaks with the lowest frequencies are well distinguished from the power spectrum. The nonlinearity renders its effect in the phonon peak broadening.  Interestingly, we found that the phonon peak of the \fpuab{} lattice is broader than the one of the \fpub{} lattice, which indicates that the asymmetric potentials provide stronger phonon-phonon interaction compared to the symmetric ones.

\section{The \fpub{} lattice}
In the following, we will give a detailed study of the \fpub{} lattice by using our variational approach. Being a special case of the \fpuab{} lattice, i.e., $\alpha=0$, it has a symmetric inter-particle potential.

\subsection{$P=0$}
The existing \QH{} theories of this model all focus on this particular case
\cite{Alabiso.95.JSP,Lepri.98.PRE,Gershgorin.05.PRL,Li.06.EL,Gershgorin.07.PRE,
He.08.PRE,Li.13.PRE}. In this part, we
not only present the comparison between our approach and MD results, but also
point out the connections between our approach and some of the existing
\QH{} theories. The first observation is $d_U=d_L=0$ [c.f., Eqs. (\ref{eq:d2}) and (\ref{eq:d1})] by noting that the potential of the \fpub{} lattice is an even function of $\delta$. Then in the following, we only need to concern about the parameter $K$.

Firstly, we focus on the \UH{} system. The corresponding parameter $K_U$ [Eq. (\ref{eq:k2})] can be calculated to yield
\begin{equation}
K^2_U-K_U-3\beta_T^{-1}=0,
\end{equation}
which has a positive solution
\begin{equation}\label{eq:b_upper}
K_U=\frac{1}{2}(1+\sqrt{1+12\beta_T^{-1}}).
\end{equation}
This result coincides with the prediction of the so called
self-consistent phonon theory (SCPT) \cite{He.08.PRE} for the \fpub{} lattice, since the SCPT based on the GB inequality Eq. (\ref{eq:gb}) as well.

Then we turn to the \LH{} system. The parameter $K_L$ [Eq. (\ref{eq:k1})] reduces to the simple form
\begin{equation}
K_L ~=~ \frac{1}{\beta_T \langle
\delta^2\rangle_{\rho_s^c}}.\label{eq:b_lower}
\end{equation}
The corresponding sound velocity $\sqrt{K_L}$ is exactly the same as that predicted by the effective phonon theory (EPT) \cite{Li.06.EL,Li.10.PRL} based on the generalized equipartition theorem \cite{Huang.87.NULL} as well as the nonlinear fluctuating hydrodynamics (\NFH{}) using the hydrodynamic approximation \cite{Spohn.14.JSP,Mendl.13.PRL,Das.14.PRE}, although the \NFH{} is not an effective theory for phonons.

\begin{figure}[tbh]
  \centering
  \includegraphics[width=0.9\columnwidth]{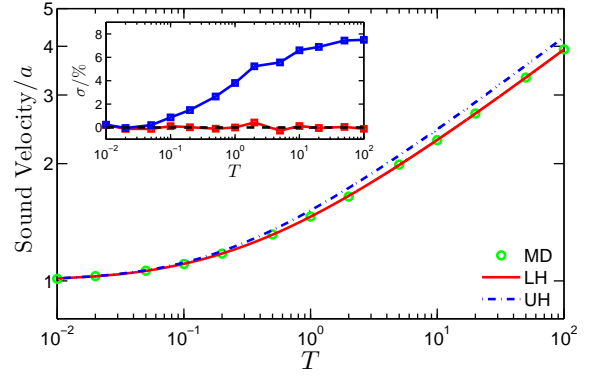}
\caption{Sound velocity as a function of temperature for the 1D \fpub{} lattice with zero pressure. The circles are MD results obtained from the power spectrum, the solid line the prediction of the \LH{} system
($\sqrt{K_L}$ [Eq. (\ref{eq:b_lower})]), and the dashed-dotted line the prediction of the \UH{} system
($\sqrt{K_U}$ [Eq. (\ref{eq:b_upper})]). The inset presents the relative deviation of the sound velocity: $\sigma=(c_{s}^{\text{TH}}-c_{s}^{\text{MD}})/c_s^{\text{MD}}\cdot 100\%$ with $c_{s}^{\text{TH}}$ and $c_{s}^{\text{MD}}$ the theoretical result and the MD result of the sound velocity, respectively. The squares denote $c_{s}^{\text{TH}}=\sqrt{K_L}$ [Eq. (\ref{eq:b_lower})] and $\sqrt{K_U}$ [Eq. (\ref{eq:b_upper})] in red and blue, respectively.}
\label{fig:bfpuv}
\end{figure}

Results of the sound velocity are illustrated in Fig. \ref{fig:bfpuv}. The
excellent agreement between the \LH{} system's predictions and MD results is
obvious. While for the \UH{} system, the deviation from MD results becomes
larger and larger as the temperature increases. The reason behind this fact is clear, since high order nonlinear terms ignored by the \UH{} system become important at high temperature. Hence the \LH{} system can describe \AP{}s in the \fpub{} lattices without pressure.

\subsection{$P\neq 0$}
\begin{figure}[tbh]
  \centering
  \includegraphics[width=0.9\columnwidth]{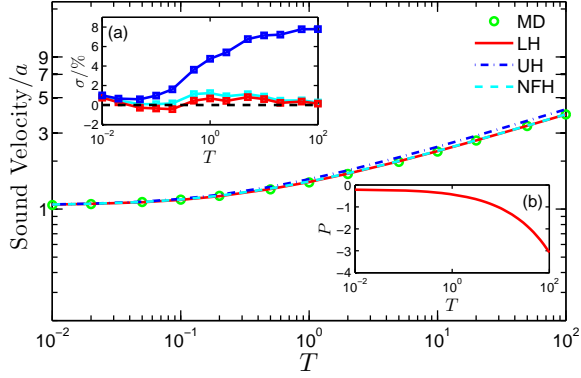}
\caption{Sound velocity as a function of temperature for the 1D \fpub{} lattice with nonzero pressure. The circles are MD results obtained from
the power spectrum, the solid line the prediction of
\LH{} system ($\sqrt{K_L}$ [Eq. (\ref{eq:k1})]), the dashed line the prediction of the \NFH{} ($\sqrt{K_{NFH}}$ [Eq. (\ref{eq:spohn1})]),
and the dashed-dotted line the prediction of the \UH{} system ($\sqrt{K_U}$ [Eq. (\ref{eq:k2})]). Inset (a) presents the relative deviation of the sound velocity: $\sigma=(c_{s}^{\text{TH}}-c_{s}^{\text{MD}})/c_s^{\text{MD}}\cdot 100\%$ with $c_{s}^{\text{TH}}$ and $c_{s}^{\text{MD}}$ the theoretical result and the MD result of the sound velocity, respectively. The squares denote $c_{s}^{\text{TH}}=\sqrt{K_L}$ [Eq. (\ref{eq:k1})], $\sqrt{K_U}$ [Eq. (\ref{eq:k1})], and $\sqrt{K_{NFH}}$ [Eq. (\ref{eq:spohn1})] in red, blue, and cyan, respectively. Inset (b) shows pressure $P$ as a function of temperature $T$ for 1D \fpub{} lattice with $\bar{L}=1.2N$.}
\label{fig:bfpuvp}
\end{figure}

Now we begin to apply our variational approach to the \fpub{} lattice with pressure. The method used to obtain nonzero pressure has been discussed in Sec. \ref{sec:md}. For simplicity, but without loss of generality, here we only study a specific situation, i.e., the lattice length is fixed to be $1.2N$. Since $\bar{L}=N(1+\left\langle \delta\right\rangle_{\rho_s})$, we have $\left\langle \delta\right\rangle_{\rho_s}=0.2$.
Therefore, the pressure $P$ can be obtained by solving
\begin{equation}
\left\langle\delta\right\rangle_{\rho_s}= \frac{\int \delta e^{-\beta_T(V+P\delta)}d\delta }{\int e^{-\beta_T(V+P\delta)}d\delta}=0.2.
\end{equation}
The pressure is temperature dependent, which is plotted in the inset (b) of Fig. \ref{fig:bfpuvp}. Since the lattice is slightly elongated, the pressure should be negative.

With those values of pressure, we can calculate the averages in Eq. (\ref{eq:k1}) for
$K_L$ numerically and thus obtain the value of $K_L$.  As for $K_U$, we insert
the values of pressure into Eq. (\ref{eq:d2}) and Eq. (\ref{eq:k2}), then solve the coupled
equations together to get the value of $K_U$. The sound velocity of the \LH{} system and the \UH{} system can be obtained
by $\sqrt{K_L}a$ and $\sqrt{K_U}a$, respectively. We present results of the sound velocity in Fig. \ref{fig:bfpuvp}. It is clearly shows that the \LH{} system still gives better results than the
\UH{} system, a significant deviation exists between the latter and MD results. So the \LH{} system can also describe \AP{}s in the \fpub{} lattices with pressure.

The prediction of the sound velocity given by the \NFH{} is also shown in Fig. \ref{fig:bfpuvp}. It takes a complex form
\begin{eqnarray}\label{eq:spohn1}
c_s/a &=& \left(\frac{\frac{1}{2}\beta_T^{-2}+\left\langle
V+P\delta ; V+P\delta\right\rangle}{\beta_T\left[\left\langle \delta ;
\delta\right\rangle\langle V ;
V\rangle-\langle \delta ;
V\rangle^2\right]
+\frac{1}{2\beta_T}\langle \delta ; \delta\rangle}\right)^{1/2}\nonumber\\
&\equiv& \sqrt{K_{NFH}},
\end{eqnarray}
in which all the $\langle\cdot\rangle$ is short for $\langle\cdot\rangle_{\rho_s}$, i.e., averages under the single site probability measure Eq. (\ref{eq:single_rho}),
and $\left\langle
A;B\right\rangle=\left\langle AB\right\rangle-\left\langle
A\right\rangle\left\langle B\right\rangle$. Noticing the EPT and the SCPT are unable to deal with systems with pressure, they become invalid in this case.

\section{The \fpuab{} lattice}
\begin{figure}[tbh]
\includegraphics[width=0.85\columnwidth]{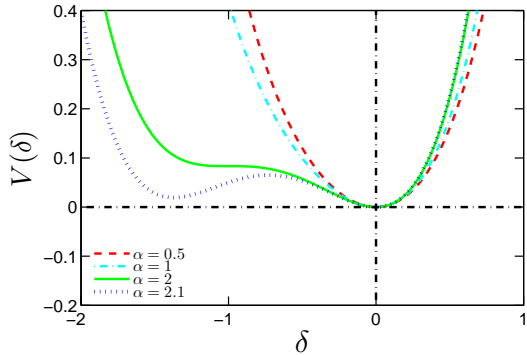}
\caption{Potential $V(\delta)$ of the 1D \fpuab{} lattice with varying $\alpha$.}
\label{fig:afpupo}
\end{figure}

When a cubic term is added to the \fpub{} potential, we get the asymmetric \fpuab{} potential [Eq. (\ref{eq:abp})]. The potential with different $\alpha$ is plotted in Fig. \ref{fig:afpupo}. It is clearly seen that the potential becomes more asymmetric as $\alpha$ increases. When $\alpha$ exceeds 2, $dV/d\delta=0$ begins to have two solutions, the potential tends to become a double-well one which is out of the scope of the present investigation, we can see that from the form of the potential with $\alpha=2.1$ in the figure.
The inherent asymmetry can induce thermal expansion, which has been shown to
play a significant role in heat conduction
\cite{Zhong.12.PRE,Wang.13.PRE,Savin.14.PRE,Das.14.JSP}. In the following, we will study this lattice by using our variational approach.

\subsection{$P=0$}
We first apply the \UH{} system to the \fpuab{} lattice. The corresponding parameters $d_U$ and $K_U$ now satisfy
\begin{eqnarray}
&&\frac{2d_U^3+d_U^4}{4}+\frac{1}{\beta_TK_U}(d_U+\alpha d_U^2+d_U^3)\nonumber\\
&&+\frac{1}{\beta^2_TK_U^2}(\alpha+3d_U)~=~0,\label{abud}\\
&&\frac{1}{\beta_T}-\frac{1}{\beta_TK_U}(1+2\alpha d_U+3d_U^2)-\frac{3}{\beta_T^2K_U^2}~=~0.\label{eq:abuk}
\end{eqnarray}
We note that $d_U$ is no longer zero due to the asymmetric potential $V$. The two equations are coupled, we should solve them together to get the value of $K_U$.

We then utilize the \LH{} system to investigate the lattice. The parameters of the \LH{} system are given by
\begin{eqnarray}
d_L &=& \left\langle\ \delta\ \right\rangle_{\rho_s^c},\\
K_L &=& \frac{1}{\beta_T \left[\langle \delta^2\rangle_{\rho_s^c}-(\langle
\delta\rangle_{\rho_s^c})^2\right]}.\label{eq:ablk}
\end{eqnarray}
Similarly, $d_L$
is nonzero for an asymmetric potential. Note that the EPT still predicts the effective force constant as \cite{Li.06.EL}
\begin{equation}
\label{eq:ept}
K_{EPT} ~=~ \frac{1}{\beta_T \langle \delta^2\rangle_{\rho_s^c}}
\end{equation}
with $V$ now the \fpuab{} potential Eq. (\ref{eq:abp}). It is evident that the denominator of $K_L$ and $K_{EPT}$ take the form of the variance and the second moment of $\delta$, respectively. This discrepancy results from the asymmetry of the potential, since $\langle\delta\rangle_{\rho_s^c}$ vanishes for symmetric potentials. As can be seen later, such a correction improves the accuracy of the results significantly.

\begin{figure}[tbh]
  \centering
  \includegraphics[width=0.9\columnwidth]{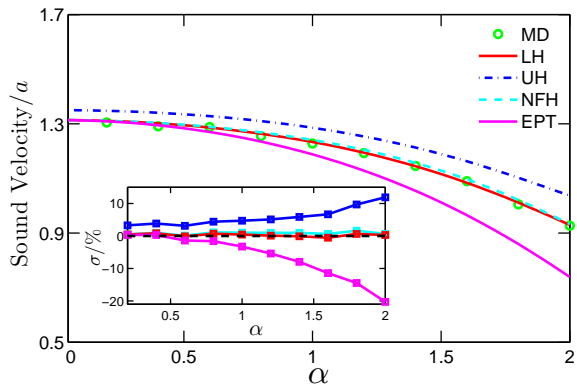}
\caption{Sound velocity as a function of $\alpha$ for the 1D \fpuab{} lattice
with $T=0.5$ and zero pressure. The circles are MD results obtained from the power spectrum, the dashed line the prediction of the \NFH{}
($\sqrt{K_{NFH}}$ with $P=0$ [Eq. (\ref{eq:spohn1})]), the dashed-dotted line the prediction of the \UH{} system
($\sqrt{K_U}$ [Eq. (\ref{eq:abuk})]), the red solid line the prediction of the \LH{} system
($\sqrt{K_L}$ [Eq. (\ref{eq:ablk})]), and the maroon solid line the prediction of the EPT
($\sqrt{K_{EPT}}$ [Eq. (\ref{eq:ept})]). The inset presents the relative deviation of the sound velocity:
$\sigma=(c_{s}^{\text{TH}}-c_{s}^{\text{MD}})/c_s^{\text{MD}}\cdot 100\%$ with $c_{s}^{\text{TH}}$ and $c_{s}^{\text{MD}}$ the theoretical result and the MD result of the sound velocity, respectively. The squares denote $c_{s}^{\text{TH}}=\sqrt{K_L}$ [Eq. (\ref{eq:ablk})], $\sqrt{K_U}$ [Eq. (\ref{eq:abuk})], $\sqrt{K_{NFH}}$ with $P=0$ [Eq. (\ref{eq:spohn1})], and $\sqrt{K_{EPT}}$ [Eq. (\ref{eq:ept})] in red, blue, cyan, and maroon, respectively.}
\label{fig:abfpuv}
\end{figure}

We consider lattices at a fixed temperature $T=0.5$ and vary $\alpha$ from 0.2
to 2. Results of the sound velocity are illustrated in
Fig. \ref{fig:abfpuv}.
The deviation of predictions of the EPT from the measured value
is clearly revealed as $\alpha$ increasing. This demonstrates the invalidity of the EPT for the strong asymmetric cases. While the \LH{} system gives accurate results no matter how large the asymmetry is. It thus shows the
significance of the correction term in the denominator of $K_L$ [Eq. (\ref{eq:ablk})]. Still, we observe that the \LH{} system gives much better results than the \UH{} system. So we can regard the \LH{} system as an effective description of \AP{}s in the \fpuab{} lattice without pressure.  Meanwhile, an agreement between the \NFH{} and the \LH{} system renders a fact that hydrodynamic approximations can also be applied to long-wavelength \AP{}s in lattices with asymmetric potentials.

\subsection{$P\neq0$}
\begin{figure}[tbh]
  \centering
  \includegraphics[width=0.9\columnwidth]{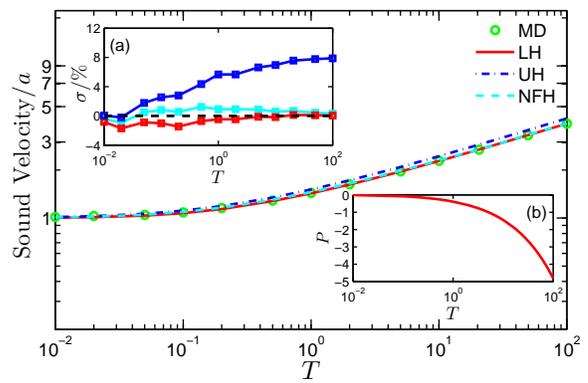}
\caption{Sound velocity as a function of temperature for the 1D \fpuab{} lattice
with $\alpha=1$ and nonzero pressure. The circles are MD results obtained
from the power spectrum, the solid line the prediction of
\LH{} system ($\sqrt{K_L}$ [Eq. (\ref{eq:k1})]), the dashed line the prediction of the \NFH{} ($\sqrt{K_{NFH}}$ [Eq. (\ref{eq:spohn1})]),
and the dashed-dotted line the prediction of the \UH{} system ($\sqrt{K_U}$ [Eq. (\ref{eq:k2})]). Inset (a) presents the relative deviation of the sound velocity: $\sigma=(c_{s}^{\text{TH}}-c_{s}^{\text{MD}})/c_s^{\text{MD}}\cdot 100\%$ with $c_{s}^{\text{TH}}$ and $c_{s}^{\text{MD}}$ the theoretical result and the MD result of the sound velocity, respectively. The squares denote $c_{s}^{\text{TH}}=\sqrt{K_L}$ [Eq. (\ref{eq:k1})], $\sqrt{K_U}$ [Eq. (\ref{eq:k1})], and $\sqrt{K_{NFH}}$ [Eq. (\ref{eq:spohn1})] in red, blue, and cyan, respectively. Inset (b) shows pressure $P$ as a function of the temperature $T$ for the \fpuab{} lattice with $\alpha=1$ at its natural length.}
\label{fig:abfpuvp}
\end{figure}

Now we consider the \fpuab{} lattice with pressure. For simplicity, We only investigate the system at its natural length. The value of the pressure can be obtained by solving
\begin{equation}
  \left\langle\delta\right\rangle_{\rho_s}= \frac{\int \delta e^{-\beta_T(V+P\delta)} d\delta}{\int e^{-\beta_T(V+P\delta)} d\delta} =0.
\end{equation}
Results of pressure as a function
of temperature are plotted in the inset (b) of Fig. \ref{fig:abfpuvp}. The absolute value of pressure
increases as the temperature increases because of the inter-particle interaction
is stronger. At the low temperature regime where thermal excitations dwell the
very bottom of the potential, particles can not feel the asymmetry of the
potential strongly, so the pressure approaches zero as the temperature tends to
zero.

With these values, $K_L$ can be easily calculated from Eq. (\ref{eq:k1}).
The value of $K_U$ can be obtained by solving Eqs. (\ref{eq:d2}) and (\ref{eq:k2}) together.
Since the existing \QH{} theories fail in this model, only our results and the
\NFH{}'s will be shown.

The result for the case $\alpha=1$ is illustrated in Fig. \ref{fig:abfpuvp}.
From this figure, we see that the \LH{} system works better than the \UH{} system as usual.

\begin{figure}[tbh]
  \centering
  \includegraphics[width=0.9\columnwidth]{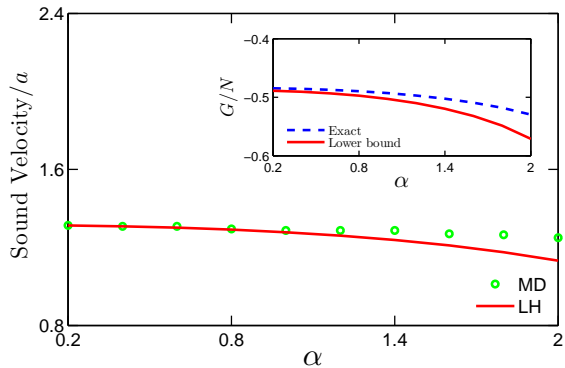}
\caption{Sound velocity as a function of $\alpha$ for the 1D FPU-$\alpha\beta$
lattice with $T=0.5$ and nonzero pressure. The circles are MD results
obtained from the power spectrum, the solid line the
prediction of $\sqrt{K_L}$ [Eq. (\ref{eq:k1})]. The inset presents the
Gibbs free energy per particle. The dashed line is the exact Gibbs free energy $g$ according to Eq. (\ref{eq:g_exact}), the solid line is the prediction of the lower bound according to Eq. (\ref{eq:low}).}
\label{fig:abfpuva}
\end{figure}

We further investigate the \fpuab{} lattices with varying $\alpha$ by using the \LH{} system. The temperature $T$ is fixed to be $0.5$. From the Fig. \ref{fig:abfpuva}, it is apparent that the discrepancy between the theoretical prediction and MD results tends to increase as $\alpha$ increases.
One possible reason is that the Gibbs free energy given by the lower bound depart from the exact result significantly at large $\alpha$ (we can see that from the inset in Fig. \ref{fig:abfpuva}). This fact indicates that the first cumulant approximation we adopted in deriving the inequalities is not enough. Higher order contributions to the Gibbs free energy must be considered in order to improve this method \cite{Mansoori.69.JCP,Mansoori.70.JCP}.

\section{Summary}
In summary, we have presented a variational approach to study renormalized phonons in momentum conserving nonlinear lattices. Specifically, we obtain two optimal harmonic reference systems, namely, the \LH{} system and the \UH{}
system, by optimizing the two bounds of the Gibbs free energy of the nonlinear system. These optimal systems which determine the properties of renormalized phonons can be regarded as optimal harmonic approximations of the nonlinear system.

This method has been applied to lattices with either symmetric or asymmetric potentials, with or without pressure. For lattices with symmetric potentials, such as the \fpub{} lattice, our variational approach gives two optimal and symmetrical reference harmonic lattices. In the zero pressure case, it is very interesting to find that the sound velocity derived from the \LH{} system reproduce the previous theoretical results from the effective phonon theory and nonlinear fluctuating hydrodynamic theory, and the sound velocity derived from the \UH{} system recover the previous theoretical results from a self-consistent approach.

For lattices with asymmetric potentials, such as the \fpuab{} lattice where the existing effective phonon theory fails to predict the sound velocity, our variational approach can also give accurate predictions. In particular, our approach reveals the reason why the effective phonon theory cannot predict the correct sound velocity.

In all the cases, the \LH{} system works better than the \UH{} system, since the ensemble averages in the former are evaluated under the probability measure of the nonlinear system, it simultaneously takes nonlinear contributions into account compared with the latter. Therefore, the \LH{} system can be treated as an effective theory of renormalized phonons in various momentum conserving nonlinear lattices. A deviation between the prediction of our variational approach and the MD results has also been found for stronger asymmetry and nonzero pressure situation where the underlying mechanism needs further investigations.

Compared with the existing \QH{} theories, this approach can be used to investigate nonlinear systems with asymmetric potentials. We owe this ability to two aspects. One is the choice of an parameterized asymmetric harmonic potential with a parameter $d$ which can quantifies the degree of asymmetry of the potentials. The other is the consideration
of the Gibbs free energy instead of the Helmholtz free energy, which enables us to deal with systems with pressure.

We also found that the \NFH{} gives satisfactory predictions of sound velocity in all those cases. The theory focuses on a hydrodynamic description of nonlinear lattices, not a construction of an effective theory of the renormalized phonons in nonlinear lattices. Those agreements means that the hydrodynamic approximation really captures essential features of the systems in the long-wavelength limit. However, for systems without momentum conservation, the concept of sound velocity is invalid, the \NFH{} may not be able to give the information of the renormalized phonons, although it can still describe correlation functions of those systems \cite{Spohn.14.A}. While our variational approach can be uesd to investigate phonon band gap and phonon dispersion of those systems \cite{Liu.14.NULL}.

\begin{acknowledgments}
We acknowledge the financial supports from the National Nature Science Foundation of China with Grant No. 11334007 (N.Li and B.Li), the National Basic Research Program of China(2012CB921401) and the Program for New Century Excellent Talents of the Ministry of Education of China with Grant No. NCET-12-0409.
\end{acknowledgments}


\begin{thebibliography}{60}
\bibitem{Lepri.03.PR}S. Lepri, R. Livi and A. Politi, Phys. Rep. {\bf377}, 1 (2003).
\bibitem{Dhar.08.AP}A. Dhar, Adv. Phys. {\bf57}, 457 (2008).
\bibitem{Liu.12.EPJB}S. Liu, X. Xu, R. Xie, G. Zhang, and B. Li, Eur. Phys. J. B {\bf85}, 337 (2012).
\bibitem{Alabiso.95.JSP}C. Alabiso, M. Casartelli, and P. Marenzoni, J. Stat. Phys. {\bf79}, 451 (1995).
\bibitem{Lepri.98.PRE}S. Lepri, Phys. Rev. E {\bf58}, 7165 (1998).
\bibitem{Alabiso.01.JPA}C. Alabiso and M. Casartelli, J. Phys. A {\bf34}, 1223 (2001).
\bibitem{Gershgorin.05.PRL}B. Gershgorin, Y. V. Lvov, and D. Cai, Phys. Rev. Lett. {\bf95}, 264302 (2005).
\bibitem{Li.06.EL}N. Li, P. Tong, and B. Li, Europhys. Lett. {\bf75}, 49 (2006).
\bibitem{Gershgorin.07.PRE}B. Gershgorin, Y. V. Lvov, and D. Cai, Phys. Rev. E {\bf75}, 046603(2007).
\bibitem{He.08.PRE}D. He, S. Buyukdagli, and B. Hu, Phys. Rev. E {\bf78}, 061103(2008).
\bibitem{Li.13.PRE}N. Li and B. Li, Phys. Rev. E {\bf87}, 042125 (2013).
\bibitem{Li.10.PRL}N. Li, B. Li, and S. Flach, Phys. Rev. Lett. {\bf105}, 054102 (2010).
\bibitem{Li.07.EL}N. Li and B. Li, Europhys. Lett. {\bf78}, 34001 (2007).
\bibitem{Zhang.13.A}Y. Zhang, S. Chen, J. Wang, and H. Zhao, arXiv:1301.2838(2013).
\bibitem{Liu.14.PRB}S. Liu, J. Liu, P. H\"{a}nggi, C. Wu, and B. Li, Phys. Rev. B {\bf90},174304(2014).
\bibitem{Girardeau.07.NULL}M. D. Girardeau and R. M. Mazo, Adv. Chem. Phys. {\bf24}, 187(1973).
\bibitem{Fermi.55.NULL}F. Fermi, J. Pasta, S. Ulam, and M. Tsingou, studies of nonlinear problems I, Los Alamos preprint LA-1940, 1955.

\bibitem{Brown.58.MP}W. B. Brown, Mol. Phys. {\bf1}, 68(1958).

\bibitem{Landau.80.NULL}L. D. Landau and E. M. Lifshitz, {\it Statistical Physics, Part I} (Pergamon, Oxford, 1980).

\bibitem{Decoster.04.JPA}A. Decoster, J. Phys. A: Math. Gen. {\bf37}, 9051 (2004).
\bibitem{Gibbs.10.NULL}J. W. Gibbs, {\it Elementary principles in statistical mechanics} (Longmans Green and Company, New York, 1928).

\bibitem{Feynman.82.NULL}R. P. Feynman, {\it Statistical Mechanics} (Benjamin, Massachusetts, 1982).

\bibitem{Morris.95.PRL}J. R. Morris and K. M. Ho, Phys. Rev. Lett. {\bf74}, 940 (1995).
\bibitem{Barnes.02.JCP}C. D. Barnes and D. A. Kofke, J. Chem. Phys. {\bf117}, 9111 (2002).

\bibitem{Li.12.RMP}N. Li, J. Ren, L. Wang, G. Zhang, P. H\"{a}nggi, and B. Li, Rev. Mod. Phys. {\bf84}, 1045 (2012).

\bibitem{Ford.92.PR}J. Ford, Phys. Rep. {\bf213}, 271 (1992).
\bibitem{Berman.05.C}G. P. Berman and F. M. Izrailev, Chaos {\bf15}, 015104 (2005).

\bibitem{Dugdale.54.PR}J. S. Dugdale and D. K. C. MacDonald, Phys. Rev. {\bf96}, 57(1954).
\bibitem{Spohn.14.JSP}H. Spohn, J. Stat. Phys. {\bf154}, 1191 (2014).

\bibitem{Laskar.01.CMDA}J. Laskar and P. Robutel, Celest. Mech. Dyn. Astron. {\bf80}, 39(2001).
\bibitem{Zhao.06.PRL}H. Zhao, Phys. Rev. Lett. {\bf96}, 140602 (2006).

\bibitem{Chatfield.89.NULL}C. Chatfield, {\it The Analysis of Time Series-An Introduction}(Chapman and Hall, London, 1989).

\bibitem{Huang.87.NULL}See e.g., K. Huang, {\it Introduction to Statistical Mechanics} (Taylor \& Francis, London, 2001).

\bibitem{Mendl.13.PRL}C. B. Mendl and H. Spohn, Phys. Rev. Lett. {\bf111},230601(2013).
\bibitem{Das.14.PRE}S. G. Das, A. Dhar, K. Saito, C. B. Mendl and H. Spohn, Phys. Rev. E {\bf90},012124(2014).

\bibitem{Zhong.12.PRE}Y. Zhong, Y. Zhang, J. Wang, and H. Zhao, Phys. Rev. E {\bf85}, 060102 (2012).
\bibitem{Wang.13.PRE}L. Wang, B. Hu, and B. Li, Phys. Rev. E {\bf88}, 052112 (2013).
\bibitem{Savin.14.PRE}A. V. Savin and Y. A. Kosevich, Phys. Rev. E {\bf89}, 032102(2014).
\bibitem{Das.14.JSP}S. Das, A. Dhar, and O. Narayan, J. Stat. Phys. {\bf154}, 204 (2014).
\bibitem{Mansoori.69.JCP}G. A. Mansoori and F. B. Canfield, J. Chem. Phys. {\bf51}, 4958(1969).
\bibitem{Mansoori.70.JCP}G. A. Mansoori and F. B. Canfield, J. Chem. Phys. {\bf53}, 1618(1970).

\bibitem{Spohn.14.A}H. Spohn and G. Stoltz, arXiv:1410.7896(2014).

\bibitem{Liu.14.NULL}J. Liu et.al., unpublished.
\end{thebibliography}
\end{document}